\documentclass{iopart}
\usepackage{iopams}
\usepackage{graphicx}
\usepackage{color}
\usepackage{bm}

\begin{document}

\title{Ettingshausen effect due to Majorana modes}
\author{C-Y~Hou$^{1,2}$, K~Shtengel$^{3,1,4}$, G~Refael$^{2}$ and P~M~Goldbart$^{5}$}
\address{$^{1}$ Department of Physics and Astronomy, University of California
at Riverside, Riverside, CA 92521, USA}
\address{$^{2}$ Department of Physics, California Institute of Technology,
Pasadena, CA 91125, USA}
\address{$^{3}$ Microsoft Research, Station Q, Elings Hall, University of California, Santa Barbara, CA 93106, USA}
\address{Department of Physics and Astronomy, University of California
at Riverside, Riverside, CA 92521, USA}
\address{$^{4}$Institute for Quantum Information, California Institute of Technology,
Pasadena, CA 91125, USA}
\address{$^{5}$ School of Physics, Georgia Institute of Technology, 837 State Street, Atlanta, GA 30332, USA}
\date{\today}

\begin{abstract}
The presence of Majorana zero-energy modes at vortex cores in a topological superconductor implies that each vortex carries an extra entropy $s_0$, given by $(k_{\mathrm{B}}/2)\ln 2$, that is independent of temperature. By utilizing this special property of Majorana modes, the edges of a topological superconductor can be cooled (or heated) by the motion of the vortices across the edges. As vortices flow in the transverse direction with respect to an external imposed supercurrent, due to the Lorentz force, a thermoelectric effect analogous to the Ettingshausen effect is expected to occur between opposing edges. We propose an experiment to observe this thermoelectric effect, which could directly probe the intrinsic entropy of Majorana zero-energy modes.
\end{abstract}
\submitto{\NJP}

\section{Introduction}
\label{sec:intro}

The search for Majorana modes in condensed matter---a subject of intense experimental effort---is driven in large part by the expectation that whenever such fermions appear as zero-energy modes bound to either vortices~\cite{Volovik1999,Read2000} or end points of superconducting quantum wires~\cite{Kitaev2001,Lutchyn2010a,Oreg2010} they are characterized by non-Abelian braiding statistics~\cite{Ivanov2001,Alicea2011a,Clarke2011a}. Such particles could, as a result, be utilized for quantum information processing~\cite{Nayak2008}.

A number of condensed matter systems, most notably the $\nu=5/2$ fractional quantum Hall state~\cite{Moore1991} and chiral $p$-wave superconductors~\cite{Volovik1999,Read2000}, are expected to host such non-Abelian quasiparticles. Alternatively, chiral superconductors supporting Majorana zero-energy modes can be fabricated as heterostructures composed of an $s$-wave superconductor and either a topological insulator~\cite{Fu2008} or a semiconductor having strong spin--orbit coupling and an additional source of Zeeman splitting~\cite{Sato2009,Sau2010a,Alicea2010a}. A modification of the latter scheme for the case of a semi-metal may remove the Zeeman splitting requirement~\cite{Chung2011}.

However, an unambiguous experimental observation of a Majorana zero-energy mode remains elusive, thus far. Following earlier proposals for the  interferometric detection of non-Abelian anyons in the $\nu=5/2$ fractional quantum Hall state~\cite{Fradkin1998,Stern2006a,Bonderson2006a}, similar ideas were put forward in the context of topological superconductivity~\cite{Fu2009b,Akhmerov2009a,Grosfeld2011b}. Another possible signature of Majorana modes would manifest itself through an unusual $4\pi$ (rather than the conventional $2\pi$) periodicity of a Josephson current as a function of the phase difference across the junction~\cite{Kitaev2001,Fu2009a}. A zero-bias tunnelling anomaly and the corresponding $2 e^2/h$ quantization of the tunnelling conductance from a single metallic channel into a Majorana zero-energy mode~\cite{Law2009,Wimmer2011} could provide another signature. While the first experimental results consistent with the latter prediction have been just reported~\cite{Mourik2012}, they cannot not address the most interesting feature of these quasiparticles, namely their non-Abelian statistics.

This brings us to another unique but less explored feature -- an intrinsic zero-temperature entropy of $s_0 = (k_{\mathrm{B}}/2) \ln 2$ per Majorana zero mode -- which is also an essential hallmark of such quasiparticles. This entropy results from the exponential growth of the ground-state degeneracy with the number of quasiparticles, a precondition for their non-Abelian statistics~\cite{Nayak2008}. Hence, a measurement of the intrinsic entropy carried by each vortex can be taken not only as an unmistakable signature of Majorana zero modes but also as an indication of their unusual statistics. It has been argued by Yang and Halperin that the presence of this zero-temperature entropy leads to an enhancement of thermopower~\cite{Yang2009}. Furthermore, it can be utilized for the adiabatic cooling of systems supporting non-Abelian anyons~\cite{Gervais2010,Yamamoto2011}.

In this paper, we show that the zero-temperature entropy carried by vortices in a topological superconductor induces a magneto-thermoelectric effect - the Ettingshausen effect. We also propose a specific setup---a heterostructure combining a topological insulator and an $s$-wave superconductor with a wide Josephson junction---that should give rise to a measurable signal for this effect under plausibly realizable experimental conditions.

\section{Thermoelectric effect}
\label{sec:thermoelectric}

Let us begin by introducing an intuitive qualitative picture of the edge thermoelectric effect in a 2D topological superconductor with broken time reversal symmetry. An edge of such a superconductor is characterized by the existence of a gapless chiral mode. (Depending on the net vorticity inside the topological superconducting region, such a mode may or may not be exactly at zero energy.) When a vortex enters the topological region, it necessarily crosses this gapless edge and changes its spectrum; as a result a quantum state is ``peeled off'' from the edge to form the Majorana zero-energy mode localized at the vortex core. This process, in turn, reduces the entropy associated with the edge modes by exactly $s_0=\frac{1}{2} k_{\mathrm{B}} \ln 2$, which is carried away by the vortex. In the reverse process, whereby a vortex moves out from the topological superconductor region, the same amount of entropy is added back into the opposing edge.

Alternatively, we could analyze the effect of a \emph{pair} of vortices entering into the topological superconductor. The advantage of this approach is that we need not concern ourselves with the details of the edge spectrum reconstruction upon the passage of each successive vortex.  Once the pair has moved deep into the topological region, the energy spectrum of the edge must return to its original form. On the other hand, both vortices now carry zero modes (provided they are sufficiently well separated). Assuming that the temperature is lower than the dimensional quantization energy inside their cores (the minigap), there are no other entropy changes associated with the vortices. On the other hand, if the vortex passage is an adiabatic process, the total entropy associated with the vortex and edge states must be conserved. The only way to reduce the entropy of the edge states in order to compensate for the entropy now carried by the two vortex zero modes is by reducing the temperature of the edge. (Only the those states that are within the `temperature window' of the edge contribute to its entropy.) Naturally, the opposite effect results from the pair of vortices leaving the topological region.

Owing to the Lorentz force, vortices may be driven across the sample via an externally imposed supercurrent. Because such vortex flow cools one and heats the other of an opposing pairs of edges, a thermoelectric effect, analogous to the Ettingshausen effect, occurs between opposing edges. As we shown below, such an effect can be quantified in terms of the ratio of the temperature difference $\Delta T$ between the opposing edges and the voltage drop $V$ in the applied current direction $\Delta T/ V = (e/k_{\mathrm{B}}) (12/\pi^2) \ln2$.

As discussed by Yang and Halperin~\cite{Yang2009}, the presence of the intrinsic entropy $s_0$ per vortex is justified only when the temperature is higher than the energy splitting of the zero-energy modes, i.e., $T \gg T_0\sim \Delta e^{-l/l_0}$. For the present work, $\Delta$ is the superconductor gap, $l_0$ is the typical size of vortex, and $l$ is the distance between vortices. Thus, in the limit of dilute vortices, $T_0$ is exponentially suppressed. At a nonzero temperature these vortices carry additional entropy, due to other minigap states~\cite{Caroli1964}. However, if the temperature is lower than the minigap, i.e.,  $k_{\mathrm{B}} T \ll \mathcal{E}_{\mathrm{mg}}$, these contributions are suppressed exponentially by $S_\mathrm{mg}\sim  |\mathcal{E}_{\mathrm{mg}}/T|  e^{-|\mathcal{E}_{\mathrm{mg}}/k_{\mathrm{B}} T|}$ and can be simply ignored (see~\ref{sec:minigap-entropy}). Therefore, the edge thermoelectric effect due to the presence of the intrinsic entropy will only be prominent in the temperature range $T_0 \ll T \ll \mathcal{E}_{\mathrm{mg}}/k_{\mathrm{B}}$.

\begin{figure}
\includegraphics[angle=0,scale=1.2]{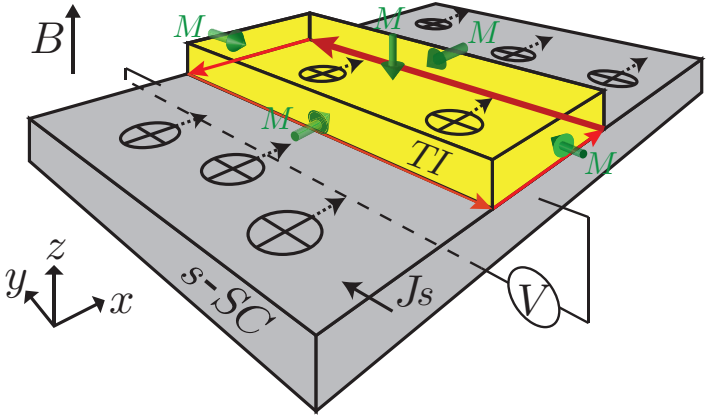}
\caption{Schematic plot of the proposed setup showing the thermoelectric mechanism for the edge state. A 3D topological insulator (colored yellow) is in contact with an $s$-wave superconductor (colored grey) from one side, and the rest of surface is coated by ferromagnetic insulator of magnetization $M$ normal to and inward from the surface indicated by the green arrow at each surface. A circulating Majorana edge state (shown in red) forms at the boundary separating the regions of superconducting pairing and magnetic gaps. A magnetic field is applied in the $z$-direction. A current driven in the $y$-direction produces a force that pushes vortices in the $x$-direction. The passage of a vortex cools the left edge and heats the edge on the opposite (i.e., right) side.
}
\label{fig:schematic}
\end{figure}

As a concrete example, we shall consider the schematic setup shown in Fig.~\ref{fig:schematic}, which illustrates the mechanism underlying this thermoelectric effect. We consider a topological insulator (TI) that interfaces with an $s$-wave superconductor, so that a superconducting pair potential is induced in the contact region of the TI via the proximity effect. This region effectively emulates a 2D topological superconductor. Each vortex in this region would have a Majorana zero-energy mode bound at its core~\cite{Fu2008}. We envision magnetically gapping the rest of the TI surface by depositing on it a ferromagnetic insulator. (We shall revisit this point and discuss more practical means of producing such a gap later on.) In this setup, a one-dimensional chiral Majorana edge state will form at the boundary of the superconducting region~\cite{Fu2009b,Akhmerov2009a}. We then imagine subjecting this region to a transverse magnetic field of strength $B>B_{\mathrm{c}1}$, which will result in a vortex density $n_v= B/\Phi_0=B/(h/2e)$ in the superconducting slab. Here, $B_{\mathrm{c}1}$ is the first critical field and $\Phi_0=h/2e$ is a superconducting flux quantum. Finally, we envision applying an external current to the superconductor, which will induce vortices to move laterally, between the two opposing edges.

If the vortices move with velocity $\mathbf{u}$ (which depends on the frictional force on moving vortices), the entropy current in the TI would be given by~\cite{Bardeen1965,Tinkham2004}
\begin{equation}
\label{eq:entropy-current-1}
 \mathbf{j}_S = s_0 n_v \mathbf{u} = s_0 \frac{2 e B}{ h } \mathbf{u} .
\end{equation}
In order to sustain a constant vortex motion, a uniform electric field $\mathbf{E}= \mathbf{B} \times \mathbf{u}$ should be applied in the direction perpendicular to both the magnetic field and the vortex motion. Hence, in terms of the applied electric field, the entropy current becomes
\begin{equation}
\label{eq:entropy-current-2}
 \mathbf{j}_S =  s_0 \frac{2 e}{ h } \mathbf{E}\times \mathbf{\hat{B}} ,
\end{equation}
where $\mathbf{\hat{B}}\equiv \mathbf{B}/\left\vert\mathbf{B}\right\vert$ is the unit vector in the direction of the magnetic field. Here, the effect of the Magnus force has been ignored, as it will not affect the conclusion given in Eq.~\eref{eq:entropy-current-2}.

To understand the heating (cooling) of edges due to the vortex flow, we first need the heat capacity per unit length $C^{\ }_V$ of the chiral Majorana edge state, which is given by
\begin{equation}
 c_{V}^{\ }= \frac{\pi^{2}}{3} k_{\mathrm{B}}^{2} T \rho^{\ }_{E_\mathrm{F}},
\end{equation}
where $\rho_{E_\mathrm{F}}= 1/(4 \pi \hbar v_{\psi})$ is the density of states at the Fermi energy, and $v_{\psi}$ is the velocity of the edge states. The energy current flowing out of/into the heated/cooled region is given approximately by
\begin{equation}
 \frac{dQ}{dt}= c^{\ }_{V} v_{\psi} \delta T = \frac{\pi}{12 \hbar} k_{\mathrm{B}}^{2} T \delta T,
\end{equation}
where $\delta T$ is the temperature variation due to heating/cooling. By balancing this energy flow with the heat added to (or removed from) the edge states due to the vortex motion crossing the edge, i.e.,
\begin{equation}
 \frac{dQ_{v}}{dt}= L T j_S= T s_0 \frac{2 e V}{ h },
\end{equation}
we obtain the result
\begin{equation}
\label{eq:temp-V}
 \delta T= \frac{6\ln2}{\pi^2} \frac{e V}{k_{\mathrm{B}}},
\end{equation}
where $L$ is the length of the heated region and $V=L |\mathbf{E}|$ is the voltage drop across the superconductor.

As the amount of entropy removed from one edge is  deposited by vortices at the other edge, the ratio of the temperature difference $\Delta T$ between the opposing edges and the voltage drop is given by
\begin{equation}
\label{eq:temp-diff}
 \frac{\Delta T}{V}= \frac{2 \delta T}{V}= \frac{12\ln2}{\pi^2} \frac{e}{k_{\mathrm{B}}} \approx 10^4 \; \mathrm{[K/V]}.
\end{equation}
Because both the temperature difference and the voltage drop directly originate from the motion of vortices, any vortex pinning should not affect this signal~\cite{Solomon1967}. As this thermoelectric response is quite substantial (i.e., $\Delta T\approx 10$ mK for $1$ $\mu$V of applied voltage), it should be possible to measure this effect provided it proves possible to measure the edge state temperature while keeping the edges isolated from the environment.

Although the setup in Fig.~\ref{fig:schematic} is useful for demonstrating the idea of the thermoelectric effect at a conceptual level, we emphasize that in reality the effect can be masked, and therefore difficult to measure, in this simple setting, due to the following reasons. Firstly, an Abrikosov vortex in an $s$-wave superconductor possesses a normal core, and thus carries entropy in addition to the contributions from the zero mode and minigap states. We note that these additional sources of entropy, although oblivious to the existence of the edge states, can build up a temperature gradient----the classical Ettingshausen effect---in the bulk of the superconductor, and hence obscure the edge thermoelectric effect~\cite{Tinkham2004,Solomon1967}.

Secondly, the motion of vortices may not be strictly perpendicular  to the direction of the applied current, instead having a Hall angle induced by the Magnus force and depending on  materials properties~\cite{Tinkham2004,Huebener1969}.

Thirdly, the motion of a vortex in a superconductor induces a non-zero resistivity that is proportional to the applied magnetic field, $\rho \propto \rho_\mathrm{n} B/B_{\mathrm{c}2}$, where $\rho_\mathrm{n}$ is the normal-state resistivity and $B_{\mathrm{c}2}$ is the second critical field strength~\cite{Bardeen1965}. Because conventional superconductor materials are characterized by a small normal-state resistivity, a superconductor with moving vortices yields a small resistivity. For a fixed voltage, the smaller the resistivity the larger the Joule heating. As a result, the superconductor can be heated considerably due to the motion of vortices.

Finally, carriers in chiral edge states in a spatially extended heating (or cooling) region may equilibrate with the environment before they can reach the thermometers. This could make it difficult to measure the temperature difference between two edge states that results from the transfer of the intrinsic entropy carried by the vortices.

\section{Wide Josephson Junction Device}

\begin{figure}
\includegraphics[angle=0,scale=1.2]{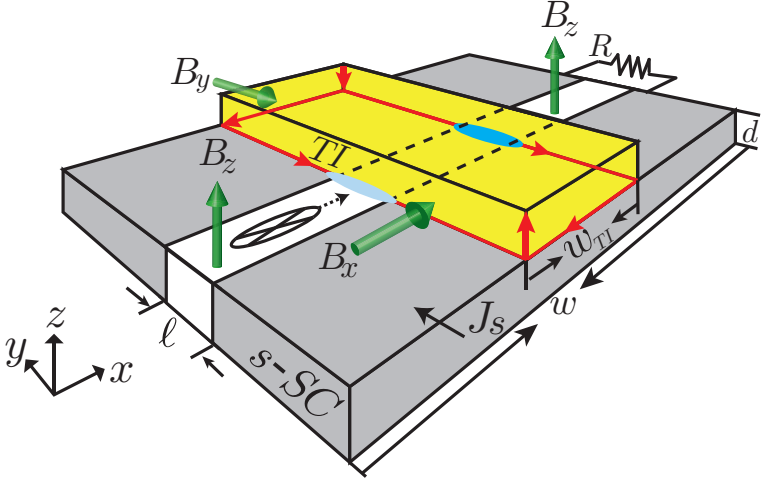}
\caption{Schematic of a setup involving a wide Josephson junction. An insulating tunnel barrier (white region) sandwiched between two $s$-wave superconductors (gray regions) forms a wide Josephson junction. The geometry of the junction is as follows: the thickness of insulator is $\ell$; the depth of junction area is $d$; the width is $w$. The width of the TI is denoted by $w^{\ }_{TI}$. The lighter and darker blue ellipses, located where edge states cross the Josephson junction, indicate cooling and heating regions, respectively. Josephson vortices are formed by applying a weak magnetic field $B_z$ in the $z$-direction. The magnetic fields are also applied, indicated by green arrows, in order to gap the surface states in both positive $x$- and negative $y$-directions. The chiral Majorana edge states appear between the superconducting and magnetic gapped regions as indicated by red perimeter with the red arrows indicating the flow directions of Majorana edge states. Between magnetic gapped surface states with opposite magnetization directions, additional edge states appear and connected to the Majorana edge states as indicated by two vertical red arrows. Here, the upper surface state of the topological insulator is not gapped as $B_z$ is very weak and is in general screened by the superconductor. An impedance-matched resistance circuit is attached at one end of junction in order to guide the vortex out of the junction. Finally, a supercurrent $J_\mathrm{s}$ is applied through the junction to drive the vortices.}
\label{fig:setup}
\end{figure}

In order to overcome the aforementioned obstacles to detecting the edge thermoelectric effect, we now propose an alternative device that utilizes Josephson junctions, as shown in Fig.~\ref{fig:setup}. A wide Josephson junction, in which Josephson vortices can propagate, is situated under a slab of TI. A constant supercurrent density $J_\mathrm{s}$ is applied across the junction (i.e., in the $y$-direction), in order to push Josephson vortices along the junction (i.e., in the $x$-direction) by means of a Lorentz force. An impedance-matched resistance circuit should be placed at one end of the junction, so that vortices are not reflect from the edge of the junction~\cite{Sakai1983,Sakai1985}. Finally, to induce a magnetic gap on the free surface of the TI, one could apply a magnetic field parallel to the interface; a chiral Majorana edge states then form at the boundary of the superconductor, as shown in Fig.~\ref{fig:setup}.

As with an Abrikosov vortex, a Josephson vortex carries a Majorana zero mode in the topological superconductor region~\cite{Grosfeld2011b}. Therefore, an analogous thermoelectric effect, caused by the mechanism described in the previous section, should occurs when dilute Josephson vortices move across the sample. A dissipative motion of such a vortex along the junction induces a voltage pulse across (due to the Josephson relation); the time-averaged voltage drop across the junction is then given by
\begin{equation}
\bar{V}= \nu \Phi_0,
\end{equation}
where $\nu$ is the passage frequency of Josephson vortices through the junction. One can subsequently show that the chiral edges emerge downstream from the junction, heated or cooled according to Eq.~\eref{eq:temp-V} with a simple substitution $V\mapsto \bar{V}$. As the cooling/heating processes only take place at or near the junction, it should be possible to probe the thermoelectric effect before the edge state equilibrates with the environment.

The advantage of dealing with wide Josephson junctions can be understood qualitatively, before providing a formal treatment of its soliton excitations (i.e., Josephson vortices). Although the superconducting phase winds by $2\pi$ around both types of vortices, a Josephson vortex has a phase core but no minigap states, whereas an Abrikosov vortex contains a normal core with minigap states~\cite{Blatter1994}. Thus, the propagation of Josephson vortices in a conventional superconductor carries no entropy, resulting in a small temperature gradient, at most. Furthermore, as the friction and pinning forces encountered by a Josephson vortex in a well-fabricated junction can be much smaller than those associated with an Abrikosov vortex moving inside a bulk superconductor, we expect much less heat dissipation due to the motion of a Josephson vortex.

The dynamics of a wide Josephson junction can be described by a sine-Gordon equation that includes damping and driving forces~\cite{McLaughlin1978}, cf.~\ref{sec:sine-Gordon-eq}:
\begin{equation}
\label{eq:sine-Gordon-All-dimensionless}
\left( \frac{\partial^2}{\partial \zeta^2} - \frac{\partial^2}{\partial \tau^2}-\alpha \frac{\partial}{\partial \tau} \right)  \varphi (\zeta,\tau)=  \sin  \varphi (\zeta,\tau) + \gamma ,
\end{equation}
where $\varphi(\zeta,\tau)$ is the position and time-dependent gauge-invariant phase difference across the Josephson junction as a function of dimensionless variables $\zeta=x/\lambda_{\mathrm{J}}$ and $\tau=\bar{c}t/\lambda_{\mathrm{J}}$. Here,
\begin{equation}
\label{eq:lambdaJ-cbar}
 \lambda^{\ }_{\mathrm{J}} \equiv \sqrt{\frac{\Phi_0}{2\pi \mu_0 (2\lambda_{\mathrm{L}} +\ell) J_{\mathrm{c}}}}, \quad \bar{c}=\sqrt{\frac{\ell}{ \epsilon \mu_0 (2\lambda_{\mathrm{L}} +\ell)} },
\end{equation}
are the Josephson penetration depth and the effective speed of light, respectively. These parameters have the meaning of a characteristic size scale and a propagation speed of a Josephson vortex along the junction; they are governed by material properties and geometric parameters of the Josephson junction: the permittivity $\epsilon$ of the insulator, its thickness $\ell$, the London penetration depth of the superconductor $\lambda_{\mathrm{L}}$, and the critical current density $J_{\mathrm{c}}$ of the Josephson junction. The damping coefficient $\alpha= \mu_0 (2\lambda_{\mathrm{L}}+\ell) \bar{c} \lambda_{\mathrm{J}}/\ell \rho$ is inversely proportional to the resistivity $\rho$ of the insulator, and the (dimensionless) driving force $\gamma= J_\mathrm{s}/J_{\mathrm{c}}$ is the ratio of the supercurrent and critical current densities.

To solve the sine-Gordon equation, one also needs to specify the boundary conditions. For a junction of width $\zeta_0$, the boundary condition for Eq.~\eref{eq:sine-Gordon-All-dimensionless} reads
\begin{equation}
\label{eq:BC-sine-Gordon}
\varphi(\zeta_0)-\varphi(0)= \frac{2\pi}{\Phi_0}B_z (2\lambda_{\mathrm{L}}+\ell) \lambda_{\mathrm{J}} \zeta_0,
\end{equation}
in the presence of applied magnetic field $B_z$. Indeed, this condition simply states that the total phase winding along the junction has to match the total flux threading through the junction.

In the limit $\alpha, \gamma \ll 1$, one can first ignore the contributions from the damping and driving forces. When the total magnetic flux in the junction area is exactly one flux quantum, the sine-Gordon equation~\eref{eq:sine-Gordon-All-dimensionless} has a soliton solution having a profile given by~\cite{Tinkham2004}
\begin{equation}
\label{eq:soliton-solu-dim}
 \varphi(x,t)= 4 \tan^{-1} \left[ \exp \left(\pm \frac{x-v t}{\lambda_{\mathrm{J}} \sqrt{1-(v/\bar{c})^2}} \right) \right],
\end{equation}
in terms of the real time and coordinate. This moving soliton, trapping exactly one flux quantum within the size $l_f=\lambda_{\mathrm{J}}  \sqrt{1-(v/\bar{c})^2}$, is a Josephson vortex. Here, the propagation velocity $v$ of the Josephson vortex is determined by the balance between the damping and driving forces and, takes the value~\cite{McLaughlin1978}
\begin{equation}
\label{eq:voetex-velocity}
  v= \bar{c}/\sqrt{ 1+ \left(4\alpha/\pi \gamma\right)^2 },
\end{equation}
which can be controlled by the ratio of damping coefficient and driving constant. In the dilute limit (i.e., vortex density $d_v < 1/\lambda_{\mathrm{J}}$), the phase profile at the Josephson junction increases monotonically, and is roughly a train of isolated Josephson vortices moving with velocity $v$. As a result, the vortex density $d_v$ matches the magnetic flux density threading through the junction, and hence can be adjusted simply by controlling the magnetic field strength $B_z$.

From energy conservation, we have that the heat dissipated per unit length by a propagating vortex is precisely the work done by the Lorentz force acting on the vortex, i.e.,
\begin{equation}
\label{eq:heat-dissi-power}
P = \Phi_0 J_\mathrm{s} v .
\end{equation}
To estimate how much energy is transferred to the edge state due to this heat dissipation, we first assume that the cross-section of the edge state is of order $\xi_s \xi$, where $\xi_s$ and $\xi$ are the coherence length of the bulk $s$-wave and the topological superconductors, respectively. Here, $\xi_s$ provides the penetration depth of the edge state in the $z$-direction, and $\xi$ is roughly the size of the edge state in the $x$-direction. Because it takes a time $\xi/v$ for a vortex to pass the edge state, we estimate that the edge state will be heated with energy
\begin{equation}
\label{eq:heat-dissi}
Q = \xi_s \xi  \Phi_0 J_\mathrm{s}.
\end{equation}
which is independent of the propagating velocity. Similarly the total heat dissipation for transferring a vortex through a junction of depth $d$ and width $w$ can be estimated as $Q_t =d w \Phi_0 J_\mathrm{s}$. Indeed, the heat dissipated by a propagating Abrikosov vortex driven by an applied supercurrent is exactly the same as that of a Josephson vortex, as given in Eq.~\eref{eq:heat-dissi-power}. However, because it requires a much smaller supercurrent to drive the vortex moving along the Josephson junction, heat dissipation becomes a much less severe issue for the wide Josephson junction device.

\section{Possible Experimental Realization}

To achieve an appreciable temperature difference, the average voltage $\bar{V}$ should be at least in the range of $0.1\sim1$ $\mu$V, which corresponds to a passage frequency of $\nu=50\sim500$ MHz. This voltage would result in a temperature difference of $\Delta T\approx 1\sim 10$ mK between the two opposing edges. With a fixed vortex velocity, the passage frequency, and hence $\bar{V}$, increases with increasing applied magnetic field $B_{z}$ (i.e., the vortex density). Hence, the temperature difference between two edges due to the Ettingshausen effect can be tuned by the magnetic field strength. To understand the issue of feasibility, let us now show that an Ettingshausen effect having a measurably large signal can be established in a wide Josephson junction device within reasonable materials parameters.

In our analysis, we assume the following wide Josephson junction geometry (see Fig.~\ref{fig:setup}): the thickness of the insulator $\ell = 2$~nm, the depth of the junction $d=5$~$\mu$m, and the width of the Josephson junction $w=0.1$~m. As a concrete example, the Josephson junction is constructed by an Al$_x$O$_y$  insulating layer of $\epsilon\approx 10 \epsilon_0$ sandwiched between a pair of $s$-wave superconductors made of Nb-Sn, and having materials properties: the superconductor pairing potential $\Delta \approx 3.4$~meV, the coherence length $\xi_{s}=3.6$~nm, and the London penetration depth $\lambda_{\mathrm{L}}=124$~nm. A Josephson junction of this type can be fabricated~\cite{Sakai1983,Sakai1985}, and is expected to have a critical current density $J_{\mathrm{c}}$ ranging from $10^5$ to $10^7$~A/m$^2$. In the present discussion, we assume $J_{\mathrm{c}}=10^6$ A/m$^2$.

Using these materials properties, together with the flux quantum $\Phi_0= 2.07\times 10^{-15}$ V$\cdot$sec, from Eq.~\eref{eq:lambdaJ-cbar} one immediately obtains
\begin{equation}
\label{eq:}
\lambda_{\mathrm{J}} \approx 32 \; \mu\mathrm{m}, \quad \bar{c}\approx 8.5 \times 10^6 \mathrm{m/s}.
\end{equation}
With the pairing potential $\Delta \approx 3.4$~meV and the critical current density $J_{\mathrm{c}}=10^6$ A/m$^2$, the tunnelling resistivity of the Josephson junction can be estimated to be $\rho\approx 2$ $\Omega\cdot$m, which corresponds to a damping coefficient $\alpha \approx 0.02$~\cite{Ambegaokar1963}.

As the vortex propagation speed has to be slower than the Fermi (edge-state) velocity of the topological insulator ($v_F\sim5\times10^5$ m/s), the wide Josephson junction should be operated in the damping-dominated regime $\gamma \ll \alpha \ll 1$. Hence, the size of a propagating Josephson vortex is $l_f\approx \lambda_{\mathrm{J}}$ since $v \ll \bar{c}$ and the width of the topological insulator should obey $w_{\mathrm{TI}}\gg \lambda_{\mathrm{J}}$. By requiring the vortex velocity $v_{\mathrm{J}}\approx 5 \times 10^4$~m/s, we obtain from Eq.~\eref{eq:voetex-velocity} the corresponding supercurrent density of $\gamma =J_\mathrm{s}/J_{\mathrm{c}} \approx 1.5 \times 10^{-4}$.

From Eq.~(\ref{eq:BC-sine-Gordon}), the density of Josephson vortices is given by $d_v= B_z(2\lambda_{\mathrm{L}}+\ell)/\Phi_0$. As the mechanism leading to the Ettingshausen effect requires a dilute vortex density, i.e.,~$d_v < 1/\lambda_{\mathrm{J}}$, we obtain the constraint $B_z< 0.25$~mT. Then, by using the passage frequency $\nu= v d_v$ and the average voltage $\bar{V}= \nu \Phi_0$, we obtain the relation
\begin{equation}
\label{eq:voltage-B}
\bar{V}= v (2 \lambda_{\mathrm{L}} +\ell) B_z.
\end{equation}
In addition, by using the aforementioned parameters, the average voltage is restricted to $\bar{V} < 3$ $\mu$V, and can be controlled directly by changing the magnetic field $B_z$. From Eq.~\eref{eq:temp-diff} and Eq.~\eref{eq:voltage-B}, we note that temperature difference is $\Delta T \approx 1\sim 10$ mK when tuning the magnetic field in the range of $B_z\approx 0.01 \sim 0.1$ mT. To enhance the strength of the Ettingshausen signal, one could introduce \emph{a set} of wide Josephson junctions, situated in parallel and separated by a distance larger than London penetration depth $\lambda_{\mathrm{L}}$. The temperature difference $\Delta T$ would then scale linearly with the number of junctions.

A superconductor in contact with a topological insulator not only produces superconductivity via proximity but also renormalizes the Fermi velocity of the surface state~\cite{Stanescu2010}. If the tunnelling rate between the superconductor and topological insulator is optimized, and the chemical potential of the topological insulator is close to the Dirac point, the induced pair-potential $\Delta_{\mathrm{TI}}\approx \Delta/2$ is about half of the bulk $s$-wave superconductor pair-potential, and the Majorana edge-state velocity $v_{\psi}\approx v_F/2$, i.e., the renormalized Fermi velocity is about half original Fermi velocity~\cite{Sau2010b}. Thus, the coherence length of the induced topological superconductor is $\xi = \hbar v_{\psi}/ \Delta \approx 25$ nm. Then, by using aforementioned material parameters, we find that the minigap of a Josephson vortex in the topological superconductor region is estimated to be
\begin{equation}
\mathcal{E}^{J}_{\rm mg} \approx \sqrt{\frac{2\hbar v_{\psi} \Delta_{\mathrm{TI} }}{\lambda_{\mathrm{J}}} } \approx 0.13\; \mathrm{meV},
\end{equation}
which equals to $1.5$ K, cf.~\ref{sec:mingap}. Taking for the operating temperature $T \sim 0.1$ K $\approx( \mathcal{E}^{J}_{\rm mg}/k_{\mathrm{B}})/15$ K, which should be readily achievable experimentally, the entropy contribution from minigap states is about $S_\mathrm{mg} \sim 0.0014\, s_0$ and can be neglected, (cf.~\ref{sec:minigap-entropy}).

Compared with the setup propagating Abrikosov vortices, the issue of heat dissipation is dramatically lessened for the wide Josephson junction setup. From Eq.~\eref{eq:heat-dissi}, we see that the heat transferred to the edge state due to a propagating Josephson vortex can be estimated as $Q \approx 10^{-28}$ J. Thus, compared with the heat added to (or moved from) the edge states due to the crossing of a vortex, $ (k_{\mathrm{B}} T/2) \ln 2\sim 5\times 10^{-25}$ J, we can ignore heating due to dissipation from vortex motion. In addition, the total heat dissipation for a vortex moving through the junction is about $Q_t\approx10^{-18}$ J/vortex, which should be drained away from the system in order to keep it at a constant temperature.

Finally, we mention that by measuring the thermopower voltage $\Delta V$ and the conductance of point-contact tunnelling into the edge state, the temperature of edge states can, in principle, be inferred via the Mott relation~\cite{Appleyard1998,Granger2009}. For the tunnelling that occurs between two normal metals (or charged 1D channels), the Mott relation reads $S=\Delta V/\Delta T= - \pi^2 k_{\mathrm{B}}^2 T (\partial \ln G/\partial \mu )/(3 e)$, where $\Delta T$ is the temperature difference between the two metals, $G$ is the conductance, and $\mu$ is the chemical potential. Although we are concerned with the tunnelling into a charge-neutral edge state in a superconductor system, we expect that the Mott relation should hold, up to an overall prefactor. We note that establishing a Mott relation for superconductors is in itself an interesting question that is worth careful examination in the future work. We should also emphasize that probing the temperatures of edge states can be a challenging experimental task. However, we envision that a setup akin to the measurement of quantum Hall edge states, as in Ref.~\cite{Granger2009}, could be a viable scheme.

\section*{Conclusion}

In summary, we have shown that the intrinsic entropy carried by vortices possessing Majorana zero-energy mode leads to an Ettingshausen effect between opposing sides of Majorana-carrying edge states. In addition, we proposed that this effect could be measured using a wide Josephson junction situated on a superconductor-topological insulator heterostructure, and we have shown that this setup should permit measurement of the Ettingshausen effect within the range of experimentally accessible parameters. At low temperature, i.e., $T\ll \mathcal{E}_{\mathrm{mg}}/k_{\mathrm{B}}$, this unique thermoelectric effect can be related to the intrinsic entropy, and thus provides a distinct probe of the non-Abelian nature of Majorana fermions. Moreover, this edge Ettingshausen effect could potentially be used as a refrigeration process for cooling small objects, such as a quantum dot.

\ack
The authors would like to thank A.~R.~Akhmerov, F.~Hassler, R.~M.~Lutchyn and C.~Nayak for helpful discussions. CYH and KS were supported in part by the DARPA-QuEST program. KS was supported in part by NSF award DMR-0748925. PMG was supported in part by NSF awards DMR-0906780 and DMR-1207026. KS, GR and PMG are grateful for the hospitality of the Aspen Center for Physics where this work was conceived. CYH would like to acknowledge the hospitality of Microsoft Station Q. GR is grateful for support from DARPA.

\appendix

\section{Entropy contribution from the minigap states}
\label{sec:minigap-entropy}

When a vortex moves from a normal superconductor into a topological superconductor region, it acquires a zero-energy quantum state. The entropy per vortex associated with such a state is $s_0=(k_{\mathrm{B}}/2) \ln 2$. In principle, the energy levels associated with other minigap states may also be affected when a vortex is moving from one region to another, in which case there will be an additional contribution to the heat transport.

The entropy carried by a fermionic state with energy $\mathcal{E}_i$ at temperature $T$ is given by
\begin{equation}
 S_i = - k_{\mathrm{B}} \left\{ p_{i} \ln (p_i)+ (1-p_i) \ln (1-p_i) \right\} ,
\end{equation}
where $p_i=1/(\exp(\mathcal{E}_i/k_\mathrm{B} T) + 1)$ is the Fermi-Dirac distribution function for chemical potential $\mu=0$. Because the entropy is extensive, the total entropy of multiple quantum states can be simply added, to give
\begin{equation}
\label{eq:entropy-sum}
S_\mathrm{mg}= \sum_{i} g_i S_i,
\end{equation}
where $g_i$ is the degeneracy. Because the entropy is suppressed exponentially by a factor $\exp(-|\mathcal{E}_{i}/k_{\mathrm{B}} T|)$ in the limit $|\mathcal{E}_{i}/k_{\mathrm{B}} T| \gg 1$, the lowest energy state (with energy $\mathcal{E}_{\mathrm{mg}}$) of the minigap states makes the most substantial contribution, and the entropy can be approximated by
\begin{equation}
S_\mathrm{mg} \approx  k_{\mathrm{B}} \left(1+ \left|\frac{ \mathcal{\mathcal{E}}_\mathrm{mg} }{k_{\mathrm{B}}T} \right| \right) \exp \left(- \left|\frac{\mathcal{E}_{\mathrm{mg}} }{k_{\mathrm{B}}T} \right| \right).
\end{equation}
Therefore, one can conclude that the entropy contribution from the minigap states can be ignored at sufficient low temperatures.

\section{Sine-Gordon equation}
\label{sec:sine-Gordon-eq}

Here, we derive the equation of motion for a wide Josephson junction in terms of the phase difference across the junction. Our derivation emphasizes the relation between the magnetic field distribution at the junction and its resulting phase, magnetic field, and current distributions inside the bulk superconductors.

Referring to the coordinates of Fig.~\ref{fig:setup}, let us consider a Josephson junction composed of an insulating layer sandwiched between two semi-infinite superconductors at areas $y \ge \ell/2$ and $y \le -\ell/2$. Hence, any $z$ dependence can be ignored. The magnetic field $\bar{B}_z(x,t)$ penetrating through the insulating area, $-\ell/2 < y < \ell/2 $, is assumed to have no $y$ dependence, and thus obeys the Maxwell equation
\begin{equation}
\label{eq:Maxwell-eq-insulator}
- \frac{\partial \bar{B}_z}{\partial x}= \mu_0 \left( J_{\mathrm{c}} \sin \varphi + J_\mathrm{s} +\frac{1}{\rho}E_{y} + \epsilon \frac{\partial E_y}{\partial t} \right),
\end{equation}
where $\varphi(x,t)$ is the gauge-invariant phase difference between two superconductors, $\mu_0$, $\epsilon$ and $\rho$ are, respectively, the permeability, permittivity and resistivity of the insulator, and $E_y$ is the electric field in the $y$-direction inside the insulator. On the right hand side, the first term is the Josephson supercurrent, the second term $J_\mathrm{s}$ is the bias supercurrent, the third term is the tunnelling current through the junction and the forth term is the displacement current. By using the Josephson relation~\cite{Tinkham2004}
\begin{equation}
 \frac{\partial \varphi}{\partial t} = \frac{2 e \ell }{\hbar} E_y,
\end{equation}
the Maxwell equation becomes
\begin{equation}
\label{eq:Maxwell-eq-insulator-1}
- \frac{\partial \bar{B}_z}{\partial x}= \mu_0\left( J_{\mathrm{c}} \sin \varphi + J_\mathrm{s} + \frac{\Phi_0 \epsilon }{ 2\pi \ell } \frac{\partial^2 \varphi}{\partial t^2} + \frac{\Phi_0}{ 2\pi \rho \ell} \frac{\partial \varphi}{\partial t}\right).
\end{equation}
Now, our goal is to associate $\bar{B}_{z}$ with $\varphi$, and thus derive an equation of motion only in terms of $\varphi(x,t)$.

For simplicity, we assume that the superconductor coherence length $\xi$ vanishes, and the healing effect due to the presence of the boundary and the magnetic field can be ignored. Because the magnetic field $\bar{B}_{z}$ at the junction effectively provides the boundary conditions for both sectors of superconductor, the magnetic field and phase distributions inside the superconductors can be solved as a boundary value problem. To further simplify matters, we take the limit $\ell \to 0$ when solving this boundary value problem. Also, we only consider the boundary value problem for the $y>0$ regions and then make use of the mirror symmetry to obtain the solution for the other half- plane.

To derive the magnetic field and phase distributions inside the superconductor, we first recall the supercurrent density
\begin{equation}
\label{eq:SC-current-density-app}
\mathbf{J}^{B}_\mathrm{s}= \frac{2 e n_\mathrm{s} \hbar}{m^*} \left( \nabla \phi - \frac{2\pi}{\Phi_0}\mathbf{A}\right),
\end{equation}
where the prefactor involves the superfluid density $n_\mathrm{s}$ and the effective mass $m^{*}$, $\phi$ is the superconductor phase, and $\mathbf{A}$ is the vector potential. When ignoring the capacitance (as we shall do), current conservation, $\nabla \cdot \mathbf{J}^{B}_\mathrm{s}=0$, leads to
\begin{equation}
\label{eq:SC-current-cons-app}
\nabla^2 \phi - \frac{2\pi}{\Phi_0}\nabla \cdot \mathbf{A} = 0.
\end{equation}
From $\nabla \times \mathbf{B}=\mu_0 \mathbf{J}^{B}_\mathrm{s}$ and $\mathbf{B}= \nabla \times \mathbf{A}$, we then obtain the Maxwell-current relation
\begin{equation}
\label{eq:SC-A-phi-relation}
 - \nabla^{2} \mathbf{A} + \nabla (\nabla\cdot \mathbf{A} ) =\mu_0 \frac{2 e n_\mathrm{s} \hbar}{m^*}  \left( \nabla \phi - \frac{2\pi}{\Phi_0}\mathbf{A}\right).
\end{equation}

To proceed, it is useful to choose a gauge. In order to read out the phase difference across the junction, it is most convenient to choose the following one:
\begin{equation}
\label{eq:gauge}
\fl
\quad
 \phi(x,y,t) = - \phi(x,-y,t),
 \qquad A_{x}(x,y,t) = -A_{x}(x,-y,t),\qquad  A_y=0,
\end{equation}
in which case the phase difference is immediately given by $\varphi(x)=2 \phi (x,y=\ell/2)$. In this gauge, current conservation \eref{eq:SC-current-cons-app} reads
\begin{equation}
\label{eq:SC-current-cons-cg}
\nabla^2 \phi - \frac{2\pi}{\Phi_0} \frac{\partial A_x }{\partial x} = 0,
\end{equation}
and Eq.~\eref{eq:SC-A-phi-relation} explicitly becomes
\begin{eqnarray}
\label{eq:SC-A-phi-relation-cg}
 -\frac{\partial^2 A_x }{\partial y^2} &=&  \frac{2 e \mu_0 n_\mathrm{s} \hbar}{m^*}\left(  \frac{\partial \phi }{\partial x} - \frac{2\pi}{\Phi_0} A_x\right),
\nonumber\\
\frac{\partial^2 A_x}{\partial x \partial y} &=&
 \frac{2 e \mu_0  n_\mathrm{s} \hbar}{m^*} \frac{\partial \phi }{\partial y} .
\end{eqnarray}
Thus by using the relation $B_z= -\partial_y A_x $, we arrive at result
\begin{equation}
\label{eq:diff-eq-Bz}
 \nabla^2 B_z-\frac{1}{\lambda_{\mathrm{L}}^2} B_z=0,
\end{equation}
together with the boundary condition $B_z(y=0)=\bar{B}_z$. Here, $\lambda_{\mathrm{L}}= \sqrt{ m^{*}/ (2 e)^2 \mu_0  n_\mathrm{s}}$ is the London penetration depth.

The differential equation \eref{eq:diff-eq-Bz} can be solved via the Green function method. The Green function with homogeneous Dirichlet boundary conditions for $y>0$ is given by
\begin{equation}
\label{eq:Green-function}
\fl
\qquad
G(x,y;x',y')
= - \int \frac{\rmd^2 k}{(2\pi)^2}  \frac{ \rme^{\rmi k_x (x-x')} \left(\rme^{\rmi k_y (y-y')}- \rme^{\rmi k_y(y+y')}\right) }{k_x^2+k_y^2+1/\lambda_{\mathrm{L}}^2 }\,.
\end{equation}
By using Green theorem, we thus have that the field distribution in the upper half-plane is given by
\begin{equation}
\label{eq:field-dis-final}
\fl
\qquad
B_z(x,y> 0)= \bar{B}(0)\, \rme^{-y/\lambda_L}
+ \int_{0^+}^{\infty} \frac{\rmd k}{ 2\pi} \, 2 \left\vert\bar{B}_z(k)\right\vert  \rme^{-\alpha_{k} y }\cos\left(k x + \theta_k \right),
\end{equation}
where $\bar{B}_z(k,t)\equiv |\bar{B}_z(k,t)| e^{i\theta_k}$ is the Fourier amplitude of the magnetic field $\bar{B}_{z}(x,t)$ at the junction, and $\alpha_k=\sqrt{ k^2+1/\lambda_{\mathrm{L}}^2 }$. We have also used the relation $\bar{B}(-k)=\bar{B}^{*}(k)$, as $\bar{B}(x,t)$ is real. The solution for $y<0$ can be inferred using mirror symmetry.

Next, observe that the component $\bar{B}_z(0)$ is the averaged magnetic field distribution, and that $\bar{B}_z(k)$ captures any spatially non-uniformity. For a uniform applied magnetic field, $\bar{B}_z(0)$ is exactly the external applied field strength and the non-vanishing $\bar{B}_z(k)$ comes solely from the non-linear current-phase response of the junction.

\begin{figure}
\includegraphics[angle=0,scale=1]{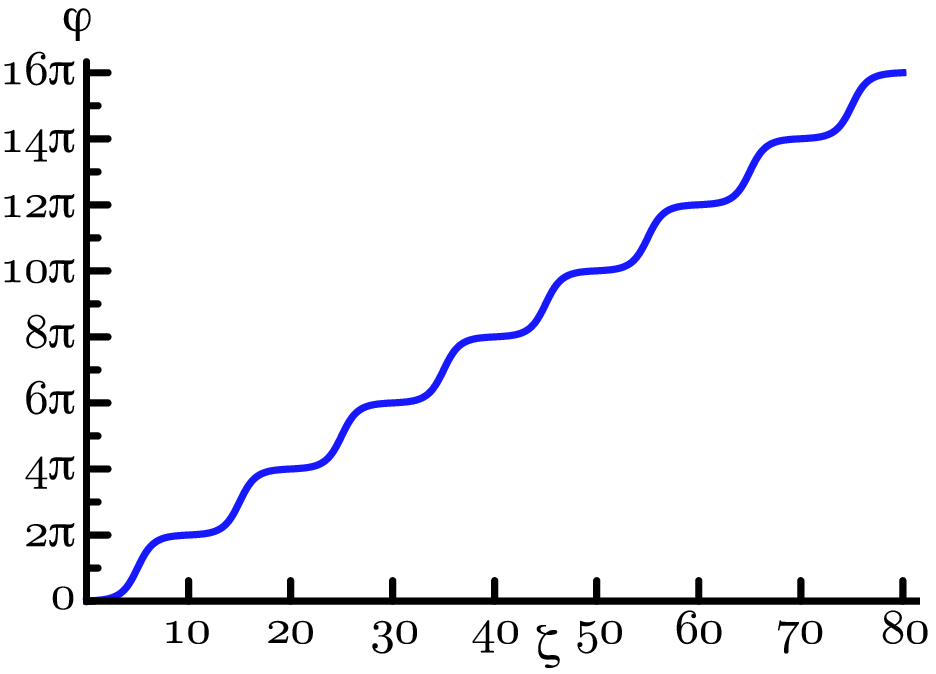}
\caption{The solution $\varphi(\zeta)$ of the sine-Gordon equation is plotted as a function of $\zeta$ for a magnetic flux density $\Phi_0/10 \lambda_{\mathrm{J}}$. As $\zeta$ is in units of $\lambda_{\mathrm{J}}$, vortices are then separated by a distance $10 \lambda_{\mathrm{J}}$ in (real) space.
}
\label{fig:Lambda-10}
\end{figure}

Upon an integration over $y$, with the boundary condition $A_x(x,y=0)=0$ imposed by the gauge choice \eref{eq:gauge}, we obtain
\begin{eqnarray}
\label{eq:vector-potential}
\fl
\qquad
 A_{x}(x,y) = \mathrm{sgn}(y) \left[ \lambda_{\mathrm{L}} \bar{B}_z(0) \left(\rme^{-\vert y\vert/\lambda_{\mathrm{L}} }-1\right)\right.
 \nonumber\\
  \left. + \int_{0^+}^{\infty} \frac{\rmd k}{ 2\pi}\, 2 \left\vert\bar{B}_z(k)\right\vert \,  \frac{\cos(k x +\theta_k)}{\alpha_k}\, \left(\rme^{-\alpha_k \vert y\vert}-1\right)\right].
\end{eqnarray}
By using Eq.~\eref{eq:SC-A-phi-relation-cg} and the  boundary condition $ J^{B}_\mathrm{s}(y\to \pm \infty)=0$, we find that the superconductor phase is given by
\begin{eqnarray}
\fl
\qquad
 \phi(x,y)= - \mathrm{sgn}(y) \frac{2\pi}{\Phi_0}\Big[ \lambda_{\mathrm{L}} \bar{B}_z(0) x
 \nonumber \\
  + \left. \int_{0^+}^{\infty} \frac{\rmd k}{ 2\pi}\,  \frac{2 \left\vert\bar{B}_z(k)\right\vert \sin(kx +\theta_k) }{k \alpha_k} \left( \lambda_{\mathrm{L}}^2 k^2 \rme^{-\alpha_k \vert y\vert} +1 \right) \right].
\end{eqnarray}
From $\varphi(x)= 2 \phi(x,0^{+})$, we thus have that the phase difference across the junction is
\begin{equation}
\label{eq:phase-diff-relation}
\fl
\qquad
 \varphi(x) = -\frac{4\pi \lambda_{\mathrm{L}}^2}{\Phi_0} \left[
  \frac {x\bar{B}_z(0)}{\lambda_{\mathrm{L}}} + \int_{0^+}^{\infty} \frac{\rmd k}{ 2\pi}\, \frac{2 \left\vert\bar{B}_z(k)\right\vert \alpha_k}{k} \sin\left(kx +\theta_k\right)  \right].
\end{equation}
This result immediately provides a way to connect the phase difference and the magnetic-field distribution at the junction.

Two important consequences can be drawn now. First, we make the observation:
\begin{equation}
 \frac{\mu_0}{2}\;  \frac{2 e n_\mathrm{s} \hbar}{m^*}\; \frac{\partial^2 \varphi(x)}{\partial x^2} = \int_{0^+}^{\infty} \frac{\rmd k}{ 2\pi}\, 2 \left\vert\bar{B}_z(k)\right\vert k \alpha_k \sin\left(kx +\theta_k\right).
\end{equation}
In the long wavelength limit, $k \ll 1/\lambda_{\mathrm{L}}$, the expansion $\alpha_k=\sqrt{k^2+1/\lambda_{\mathrm{L}}^2} \approx \frac{1}{\lambda_L} (1+ \frac{1}{2} k^2 \lambda_{\mathrm{L}}^2 +....) $ leads to the relation
\begin{equation}
 - \frac{\partial \bar{B}_{z}(x)}{\partial x} \approx  \frac{ e \mu_0 n_\mathrm{s} \hbar \lambda_L}{m^*}\; \frac{\partial^2 \varphi(x)}{\partial x^2} + \mathcal{O}\left( \frac{\partial^4 \varphi(x)}{\partial x^4}  \right).
\end{equation}
By inserting this relation into Eq.~(\ref{eq:Maxwell-eq-insulator-1}), we obtain the equation of motion
\begin{equation}
\label{eq:EOM-junction-appr}
\frac{ e \mu_0 n_\mathrm{s} \hbar \lambda_L}{ m^*}\; \frac{\partial^2 \varphi(x,t)}{\partial x^2}   = \mu_0 J_{\mathrm{c}} \sin\left( \varphi(x,t) \right) + \cdots .
\end{equation}
Then by defining the corresponding length scales $\lambda_{\mathrm{J}} = \sqrt{\frac{\Phi_0}{2\pi \mu_0 J_{\mathrm{c}} (2\lambda_{\mathrm{L}})} }$ and $\bar{c}=\sqrt{\frac{\ell}{ \epsilon \mu_0 (2\lambda_{\mathrm{L}} +\ell)} }$, and using dimensionless variables $\zeta = x/\lambda_{\mathrm{J}}$ and $\tau= \bar{c} t/\lambda_{\mathrm{J}}$, we arrive at the dimensionless sine-Gordon equation
\begin{equation}
\label{eq:sine-Gordon-dimensionless-app}
\left( \frac{\partial^2}{\partial \zeta^2} - \frac{\partial^2}{\partial \tau^2}-\alpha \frac{\partial}{\partial \tau} \right)  \varphi (\zeta,\tau)=  \sin  \varphi (\zeta,\tau) + \gamma ,
\end{equation}
Here, $\alpha$ and $\gamma$ are defined in the main text, and the Josephson penetration depth $\lambda_{\mathrm{J}}$ recovers the given in the main text provided the effect of the insulator thickness $\ell$ is taken into account.

Second, the boundary condition on the phase difference can be inferred from Eq.~\eref{eq:phase-diff-relation}. The phase consists of two contributions: a term linearly increasing with $x$ and an oscillating one. For a wide junction, i.e., $w \gg \lambda_{\mathrm{J}}$, the contribution of the oscillating term is negligible (on average) by comparison to the linear term. This translates into the condition
\begin{equation}
\label{eq:BC}
 \varphi(w)-\varphi(0)= - 2\pi \frac{\Phi_{\mathrm{ eff}}}{\Phi_0},
\end{equation}
where $\Phi_{\mathrm{eff}}=  (2 \lambda_{\mathrm{L}} ) w \bar{B}_{z}(0) $ is the effective total flux threading through the junction area. Because $\bar{B}_{z}(0)$ is the average applied magnetic field, the boundary condition for Eq.~(\ref{eq:EOM-junction-appr}) is controlable externally. Again, the effective total flux yields an additional correction with $2 \lambda_{\mathrm{L}} \to 2 \lambda_{\mathrm{L}}+\ell$ once the effect of the insulator thickness $\ell$ is taken into account.

In the limit $\alpha,\gamma \ll 1$, the phase difference profiles can be approximated by first neglecting those terms. The solution of sine-Gordon equation can be cast in term of elliptic integrals~\cite{McLaughlin1978} with the boundary condition \eref
{eq:BC}. In Fig.~\ref{fig:Lambda-10}, we plot a typical phase profile for an effective flux density of $\Phi_0/10 \lambda_{\mathrm{J}}$, as a function of the dimensionless length $\zeta$. We observe that solitons (Josephson vortices) form, separated by $10\lambda_{\mathrm{J}}$.

\section{Fermionic states bound to a Josephson vortex in a topological superconductor}
\label{sec:mingap}

Within the topological superconductor region, the fermionic states bound to a Josephson vortex can be understood via a simple edge-state coupling model~\cite{Grosfeld2011b}. The low-energy fermionic degrees of freedom along a wide Josephson junction can be described by the Hamiltonian
\begin{equation}
\label{eq:H-edge-coupling}
\fl
\qquad
H  = \rmi v_{\psi} \int \rmd x \left( \psi_{R} \partial_x \psi_{R} -\psi_{L} \partial_x \psi_{L}   \right)
+ 2 \rmi\, m \int \rmd x \cos(\varphi/2)  \psi_{R} \psi_{L} .
\end{equation}
where $\psi_{R(L)}$ is the right(left) moving Majorana fermion operator, $v_{\psi}$ is the velocity of the Majorana edge states and $m$ is the tunnelling amplitude. The first term is the free Hamiltonian of a pair of counter-propagating Majorana edge states, and the second term accounts for the tunnelling amplitude, which depends on the superconducting phase difference $\varphi$ across the junction.

The Hamiltonian (\ref{eq:H-edge-coupling}) satsifies the quantum mechanical supersymmetry that guarantees the presence of a Majorana zero-energy mode~\cite{Grosfeld2011b}. This zero-mode is responsible for the intrinsic entropy, $(k_{\mathrm{B}} /2) \ln 2$. To obtain the low-energy excitations, we first note that the the Josephson vortex profile, $\varphi= 4 \tan^{-1} \exp(x/l_f)$, leads to $\cos(\varphi/2)= -\tanh(x/l_{f})$. By linearizing the tunnelling term with the profile of $\varphi$, one can solve for the energy spectrum, to obtain
\begin{equation}
\label{eq:gap-spectrum-fluxon}
\mathcal{E}^{J}_{n} = \pm \sqrt{\frac{2\hbar v_{\psi} m}{l_f} n} \quad n=0,1,2,\ldots ,
\end{equation}
with a degeneracy $g_n=1$ for each state~\cite{Herbut2011a}. Hence, the minigap can be estimated as
\begin{equation}
\label{eq:luxon}
\mathcal{E}^{J}_{\rm mg} = \sqrt{\frac{2\hbar v_{\psi} m}{l_f} } \approx \sqrt{\frac{2\hbar v_{\psi} \Delta_{\mathrm{TI} }}{l_f} } .
\end{equation}
where we have approximate the tunnelling amplitude by $m \sim \Delta_{\mathrm{TI}}$, appropriate for a transparent junction. From Sec.~\ref{sec:minigap-entropy}, we have that the entropy contributions due to the presence of the minigap states can be estimated from this minigap.

\section*{References}
\bibliographystyle{iopart-num}
\bibliography{kirref}

\end{document}